\documentclass[openacc]{rstransa}%%%%where rstrans is the template name
\usepackage{ragged2e}
\usepackage[justification=RaggedRight,singlelinecheck=false]{caption}
\usepackage[T1]{fontenc}
\usepackage{lmodern}
\usepackage{ulem}

\newcommand{\la}{\left\langle}
\newcommand{\ra}{\right\rangle}
\newcommand{\be}{\begin{equation}}
\newcommand{\ee}{\end{equation}}
\newcommand{\bse}{\begin{subequations}}
\newcommand{\ese}{\end{subequations}}
\newcommand{\bea}{\begin{eqnarray}}
\newcommand{\eea}{\end{eqnarray}}
\newcommand{\ba}{\begin{array}}
\newcommand{\ea}{\end{array}}
\newcommand{\bi}{\begin{itemize}}
\newcommand{\ei}{\end{itemize}}
\newcommand{\ben}{\begin{enumerate}}
\newcommand{\een}{\end{enumerate}}
\titlehead{Research}
\begin{document}

\title{Structure Functions and Intermittency for Coarsening Systems}
\author{Pradeep Kumar Yadav$^{1}$, Mahendra K. Verma$^{2,3}$ and Sanjay Puri$^{4}$}
%%%%%%%%% Insert author address here
\address{$^{1}$ Department de Física de la Matèria Condensada, Universitat de Barcelona, Martí i Franqués 1, 08028 Barcelona, Spain. \\
$^{2}$ Department of Physics, Indian Institute of Technology, Kanpur -- 208016, India. \\
$^{3}$ Kotak School of Sustainability, Indian Institute of Technology, Kanpur -- 208016, India. \\
$^{4}$ School of Physical Sciences, Jawaharlal Nehru University, New Delhi -- 110067, India.}

%%%% Subject entries to be placed here %%%%
\subject{Statistical physics, fluid mechanics}
%%%% Keyword entries to be placed here %%%%
\keywords{Intermittency, domain growth, structure function}
%%%% Insert corresponding author and its email address}
\corres{Sanjay Puri \\
\email{purijnu@gmail.com}}	

%%%% Abstract text to be placed here %%%%%%%%%%%%
\begin{abstract}
In studies of turbulence, there has been extensive use of physical quantities such as {\it energy transfers} and {\it structure functions}. We examine whether these quantities can be useful in understanding problems of domain growth or coarsening, as modeled by the {\it time-dependent Ginzburg-Landau} (TDGL) equation and the {\it Cahn-Hilliard} (CH) equation. This paper has two major themes. First, we review our recent papers on energy transfers in domain growth. Second, we study structure functions and intermittency for coarsening systems. As a consequence of sharp interfaces, the structure functions scale as $S_q \sim r^{\zeta_q}$, where $r$ is the distance between two points. For the TDGL and CH models, $\zeta_q = 1$, indicating {\it anomalous scaling}.
\end{abstract}

\begin{fmtext}
\end{fmtext}
	
\maketitle
\section{Introduction}
\label{s1}

Turbulence is an ubiquitous phenomenon that arises in a wide range of natural and engineered systems, e.g., blood flow in arteries, smoke rising from chimneys, waterfalls, flows in the wake of boats, flows around aircraft wings, ocean currents, and atmospheric circulation \cite{Manneville:book:Instabilities,Manneville:book:Structures,Frisch:book,
Kolmogorov:DANS1941Dissipation,Kolmogorov:DANS1941Structure}. The physics of turbulence is very complex and depends on many factors, for example, confining walls, external fields such as gravity, forcing functions, and dissipation mechanisms. An important landmark in this field is the universal theory of turbulence due to Kolmogorov \cite{Kolmogorov:DANS1941Dissipation,Kolmogorov:DANS1941Structure}, which is often used to describe turbulent flows.

In Kolmogorov's theory for \textit{homogeneous and isotropic turbulence},  in the absence of an external field and far away from the walls,  the third-order {\it structure function} in the \textit{inertial range} is \cite{Kolmogorov:DANS1941Dissipation,Kolmogorov:DANS1941Structure}
\be 
S_3(r) = \la | ({\bf u(x+r) - u(x)}) \cdot \hat{r} |^3 \ra  = - \frac{4}{5} \epsilon r ,
\label{eq:S3_K41}
\ee
where the angular brackets represent an ensemble average. Here, ${\bf u(x)}$ and ${\bf u(x+r)}$ are the velocity fields at the positions $\bf{x}$ and $\bf{x+r}$, respectively (we have suppressed the time dependence for notational convenience). In addition, $\hat{r}$ is the unit vector along ${\bf r}$; and $\epsilon$ is the \textit{energy dissipation rate}. The above law is applicable in the \textit{inertial range}, which lies between the forcing length scale and the dissipation scale.  Many experiments and numerical simulations have validated Eq.~(\ref{eq:S3_K41})~\cite{Frisch:book}.

A simple scaling argument applied to Eq.~(\ref{eq:S3_K41}) yields~\cite{Lesieur:book:Turbulence}
\be
\la |{\bf u(x+r) - u(x)}|^2 \ra \sim r^{2/3} .
\ee
An equivalent quantity is the correlation function
\be
C(r) = \la {\bf {\bf u(x + r)} \cdot {\bf u(x)}} \ra \simeq C_1 - C_2 r^{2/3} ,
\ee
where $C_1, C_2$ are constants. This correlation function yields the famous Kolmogorov energy spectrum~\cite{Lesieur:book:Turbulence,Frisch:book}, 
\be
E_{\rm sp}(k) =\frac{1}{2} \sum_{k \le |{\bf k}| < k+\Delta k} |\hat{\bf u}({\bf k})|^2 \sim k^{-5/3} , 
\ee
where $\hat{\bf u}({\bf k})$ is the Fourier transform of ${\bf u(x)}$. The Kolmogorov law has been verified in many experiments and numerical simulations \cite{Frisch:book,Lesieur:book:Turbulence}. 

Unfortunately, despite great efforts, there is no known expression for the $q^{\rm th}$ $(q \ne 3)$ order structure function:
\be 
S_q(r) = \la | ({\bf u(x+r) - u(x)}) \cdot \hat{r} |^q \ra  .
\ee
A simple scaling argument predicts that~\cite{Lesieur:book:Turbulence,Frisch:book}
\be
S_q(r,t) \sim \epsilon^{q/3} r^{q/3}, 
\ee
which does not match with experimental observations. In this context, researchers have studied various phenomenological models, e.g., \textit{fractal model}~\cite{Frisch:JFM1978}, \textit{lognormal model}~\cite{Kolmogorov:JFM1963}, \textit{multifractal model}~\cite{Meneveau:PRL1987}, \textit{She-Leveque model}~\cite{She:PRL1994}. Among these, the She-Leveque model \cite{She:PRL1994} provides the best fit to numerical and experimental observations:
\be
S_q(r) \sim r^{\zeta_q},
\ee
where the \textit{intermittency exponent} $\zeta_q$ is
\be
\zeta_q = \frac{q}{9} + 2 \left[ 1-\left( \frac{2}{3} \right)^{q/3} \right].
\ee

In an independent set of developments, there has been great interest in the field of {\it domain growth} or {\it kinetics of phase transitions} or {\it coarsening} \cite{Puri:book_edited,Bray:AP1994}. Here, one studies the far-from-equilibrium evolution of a system which has been rendered thermodynamically unstable by a rapid change of parameters, e.g., temperature, pressure, etc. The subsequent evolution is characterized by the emergence and growth of domains enriched in different phases. In coarsening systems, large-scale structures emerge through the coalescence of smaller ones. These domain growth problems are ubiquitous and arise in all branches of science. There are many models of domain growth, ranging from phenomenological approaches such as the {\it time-dependent Ginzburg-Landau} (TDGL) and {\it Cahn-Hilliard} (CH) equations \cite{Hohenberg:RMP1977}, to lattice-based models, e.g., kinetic Ising models with Glauber or Kawasaki dynamics \cite{Puri:book_edited}. The TDGL and CH equations model the evolution of an order parameter field $\psi (\bf{x},t)$ which depends on space and time. These models reproduce the characteristic scaling laws of coarsening. In particular, the typical domain size grows as $R(t) \sim t^{\phi}$, with $\phi = 1/2$ for the TDGL equation (nonconserved kinetics) and $\phi = 1/3$ for the CH equation (conserved kinetics). Furthermore, the two-point correlation function exhibits dynamical scaling as
\begin{eqnarray}
C({\bf r}, t) &=& \la \psi ({\bf x} + {\bf r},t) \psi ({\bf x},t) \ra
\nonumber \\
&=& f\left(\frac{r}{R(t)}\right), 
\end{eqnarray}
where $f(z)$ is the scaling function. Although second-order correlations capture broad features of domain growth, higher-order structure functions are needed to reveal intermittent fluctuations. Inspired by turbulence studies, where structure functions quantify velocity correlations and scaling laws~\cite{Frisch:book}, we compute and analyze higher-order structure functions for coarsening systems governed by the TDGL and CH equations.

In isotropic turbulence, the energy cascades from large to small scales, producing the $k^{-5/3}$ spectrum through a conserved flux. In contrast, coarsening systems lack such multiscale cascades or flux conservation. However, our recent studies \cite{Verma:PRE2023_coarsening,Yadav:PRE2024} have shown that nonlinear terms in coarsening models actively dissipate fluctuations and facilitate spectral energy transfers between Fourier modes. In TDGL dynamics, these transfers eventually accumulate energy in the $k=0$ mode as the system approaches a uniform state. In CH dynamics, energy is redistributed similarly among modes, but the $k=0$ mode remains inactive due to conservation constraints. These findings reveal that, despite the absence of a traditional energy cascade, nonlinear energy transfers in coarsening systems provide valuable insights into domain growth and interface dynamics.

An important characteristic of coarsening systems is the {\it Porod law} \cite{gp82,opmpl}, which is due to scattering from sharp interfaces between domains. In analogy to turbulence, let us define the modal energy spectrum as
\be
E({\bf k},t) = \frac{1}{2} |\hat{\psi}({\bf k},t)|^2 ,
\ee
where $\hat{\psi} (\mathbf{k},t)$ is the Fourier transform of the order parameter. In the coarsening literature, this quantity is commonly referred to as the structure factor (not to be confused with the structure functions introduced earlier), and is the Fourier transform of $C({\bf r}, t)$. The structure factor is measured in scattering experiments used to study coarsening. Porod's law predicts that, for an isotropic system, the structure factor tail scales as ~\cite{Puri:book_edited,Bray:AP1994}
\be
E({\bf k}, t) \sim k^{-(D+1)},
\ee
where $D$ is the dimensionality. The corresponding result for the second-order correlation function (valid for small $r$) is
\be
C(r,t) = C_1 - C_2 r + \cdots ,
\ee
with constant $C_1$ and $C_2$.

In spite of decades of efforts, several issues in coarsening remain unresolved. For example, there are no definitive studies of higher-order correlations or structure functions for coarsening systems.  In this paper, we compute the $q^{\rm th}$ order structure function for the TDGL and CH equations. We show that
\be 
S_q(r,t) = \la |\psi({\bf x+r},t) - \psi({\bf x},t) |^q \ra \sim r , \quad \xi \ll r \ll R(t) ,
\label{sqpsi}
\ee
where $\xi$ is the thickness of the interface. The above relation is similar to that for the Burgers equation, where $S_q(r,t) \sim r$ is due to shocks \cite{Bouchaud:PRE1995,Verma:PA2000,Frisch:book}.  Note that the shocks in the Burgers equation are analogous to the domain walls in coarsening systems.

 Another important issue is the transfer of $\psi^2$ across scales. It has been argued that an inverse cascade leads to coarsening or formation of large-scales structures. However, our recent works using spectral analysis \cite{Verma:PRE2023_coarsening, Yadav:PRE2024} showed that no associated flux exists for $|\hat{\psi}({\bf k},t)|^2$ in the TDGL and CH equations. Instead, the nonlinear term enhances the diffusion of large-scale Fourier modes, which drives coarsening. In the present paper, we review these results to provide an overall picture of the utility of the turbulence framework in the context of domain growth.

This paper is organized as follows. Section~\ref{sec:ET} introduces the TDGL and CH equations, together with our framework for nonlinear energy transfers and structure functions. Section~\ref{sec:NR} presents numerical results for the structure functions. We conclude in Section~\ref{sec:conclusion}.

\section{Energy Transfers and Structure Functions for Coarsening Systems}
\label{sec:ET}

In this section, we summarize the energy transfer framework for the TDGL and CH equations. We also derive the structure functions for these systems. 

\subsection{TDGL and CH equations}

The TDGL and CH equations are formulated using the functional derivative of the Ginzburg-Landau free energy functional $\mathcal{F} [\psi]$ \cite{Puri:book_edited,Bray:AP1994}:
\begin{eqnarray}
	\frac{\partial \psi (\mathbf{x},t)}{\partial t} &=& -(-\nabla^{2})^{\alpha} \left[ \frac{\delta {\mathcal{F}}[\psi]}{\delta \psi} \right] ,
	\label{eq:models}
\end{eqnarray}
where $\alpha=0$ for the TDGL equation and $1$ for the CH equation. The order parameter $\psi({\bf x},t)$  is (a) the magnetization in the TDGL equation; and (b) the local concentration difference in a binary mixture for the CH equation. In dimensionless units, the expression for $\mathcal{F}$ is
\begin{eqnarray}
	F[\psi(\mathbf{x},t)] = \int d^D x \, \left[- \frac {\psi^2} {2}+ \frac{\psi^4}{4} + \frac {(\mathbf{\nabla} \psi)^2}{2} \right].
	\label{eq:gl}
\end{eqnarray}
The first two terms on the right-hand side of Eq.~\eqref{eq:gl} represent the bulk free energy, while the last term accounts for the penalty due to spatial variations in $\psi({\bf x},t)$. Taking the functional derivative of Eq.~\eqref{eq:gl} and substituting it into Eq.~\eqref{eq:models}, we obtain the following evolution equation:
\begin{eqnarray}
	\frac{\partial \psi(\mathbf{x},t)}{\partial t} &=& (-\nabla^2)^{\alpha}\left( \psi -\psi^{3} + \nabla^2\psi \right) .
	\label{eq:tdglch}
\end{eqnarray}
Note that $\int d{\bf x} \psi({\bf x},t)$ is conserved for the CH equation, but not for the TDGL equation.

The steady-state solutions of the TDGL and CH equations are obtained by setting $\partial / \partial t = 0$.  Both equations have the following stationary solutions in 1D:
\be
\psi(x)=0,~~\pm 1,~~\tanh \left[\pm \frac{(x- \sigma )}{\sqrt{2}} \right] , 
\label{eq:psi_asymptotic}
\ee
where $\sigma$ is the position of the kink (K) or anti-kink (AK). Additionally, $\psi(x)= 0$ and $\psi(x) = \pm 1$ are the unstable and stable solutions, respectively.
\begin{figure}[htb]
\centering  
\includegraphics[width=0.6\linewidth]{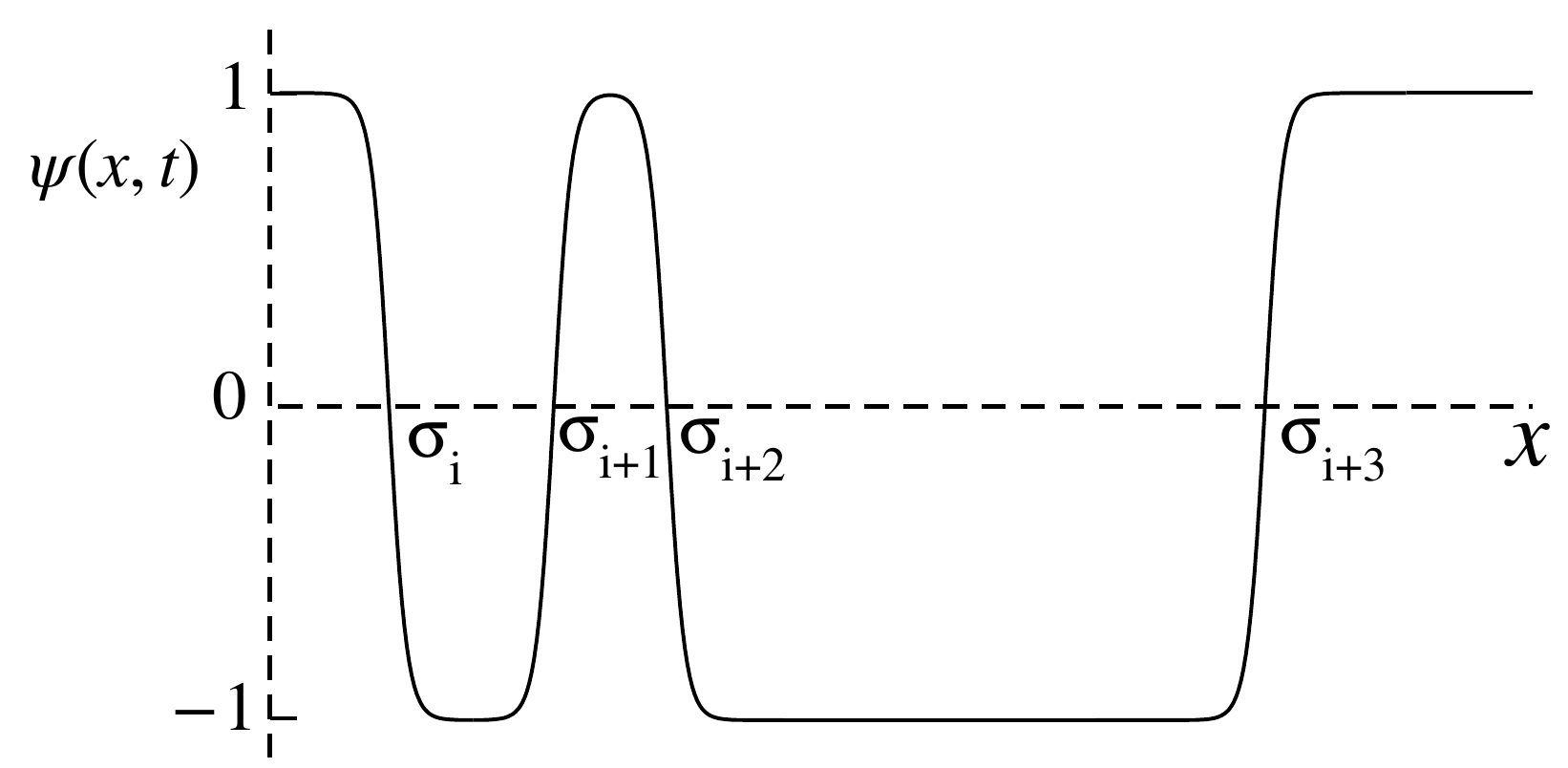}
\caption{Plot of the order parameter $\psi(x,t)$ vs. $x$, depicting a typical order parameter profile of the TDGL equation. The locations of kinks and anti-kinks are denoted by $\sigma_n$.}
\label{fig:sf1}
\end{figure}
\begin{figure}[htb]
\centering
\includegraphics[width=0.5\linewidth]{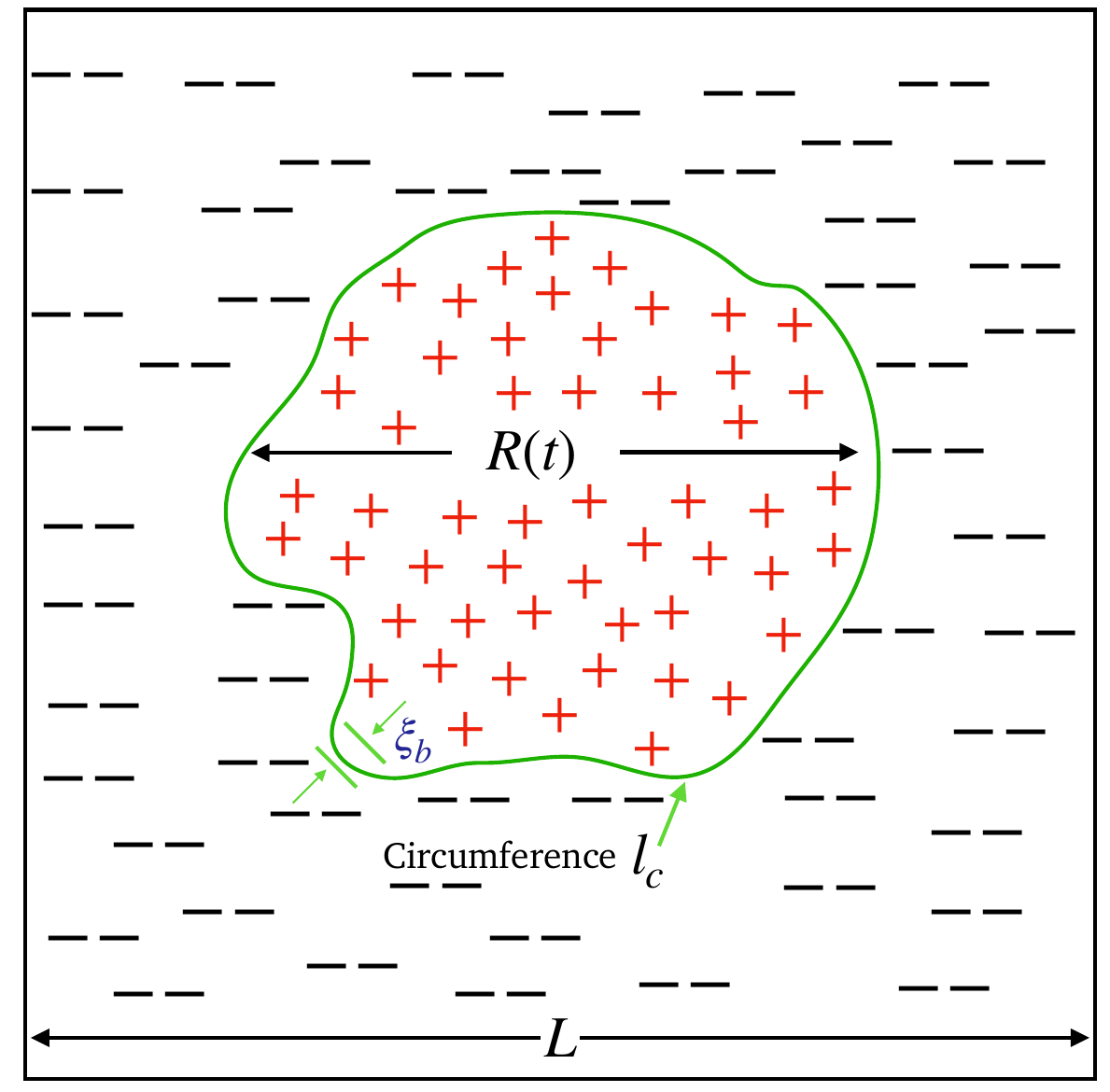}
\caption{Schematic of domain of size $R(t)$ at time $t$, with an interface of width $\xi$, in a system of linear size $L$. The $+$ region, where $\psi > 0$, is surrounded by the $-$ region, where $\psi < 0$. The circumference of the $+$ domain is denoted by $l_c$.}
\label{fig:sf2}
\end{figure}

In the next subsection, we discuss nonlinear energy transfers in the TDGL and CH equations.

\subsection{Nonlinear energy transfers}
\label{subsec:ET}

We focus on the unbiased case with $\psi({\bf x},t=0) = 0+$small fluctuations.
Equation~\eqref{eq:tdglch} in Fourier space is
\begin{eqnarray}
	\frac{\partial\hat{\psi}({\bf k},t)}{\partial t} = k^{2\alpha}  \left[\hat{\psi}({\bf k},t) - k^2 \hat{\psi}({\bf k},t) - \sum_{{\bf k}_1,{\bf k}_2,{\bf k}_3}^{'} \hat{\psi}({\bf k}_1,t) \hat{\psi}({\bf k}_2,t) \hat{\psi}({\bf k}_3,t) \right] ,
	\label{eq:phi_k}
\end{eqnarray} 
where 
\begin{equation}
	\hat{\psi}({\bf k},t) = \sum_{\bf x} e^{i {\bf k} \cdot {\bf x}} \psi({\bf x},t)   
\end{equation}
is the Fourier transform of $\psi({\bf x},t)$. The prime in the summation denotes the constraint ${\bf k} = {\bf k}_1 + {\bf k}_2 + {\bf k}_3$.  Using Eq.~(\ref{eq:phi_k}), we obtain the evolution equation for the modal energy as \cite{Verma:PRE2023_coarsening,Yadav:PRE2024}
\begin{eqnarray}
	\frac{\partial E({\bf k},t)}{\partial t}  =  k^{2\alpha} \left[2E({\bf k},t) - 2k^2 E({\bf k},t) + T({\bf k},t)\right].
	\label{eq:en}
\end{eqnarray}
Here,
\begin{equation}
	T({\bf k},t)  =   
	-\sum_{{\bf k}_1,{\bf k}_2,{\bf k}_3}^{'} \operatorname{Re}\left\{\hat{\psi}({\bf k}_1,t) \hat{\psi}({\bf k}_2,t) \hat{\psi}({\bf k}_3,t) \hat{\psi}^*({\bf k},t) \right\}.
	\label{tdgl_tkt}
\end{equation}
In Eq.~(\ref{eq:en}), (a) the term $2 k^{2\alpha} E({\bf k},t)$  acts as an energy source for the wavenumber ${\bf k}$, similar to energy injection on large scales in hydrodynamic turbulence; (b) the term $2 k^{2\alpha+2}   E({\bf k},t)$ dissipates energy; (c) the term $k^{2\alpha} T({\bf k},t)$ represents the rate of nonlinear energy transfer. Note that Eq.~(\ref{eq:en}) is not applicable to ${\bf k}=0$. 

According to Porod's law~\cite{Porod:book_chapter}, at large wavenumbers, the modal energy is
\be
E({\bf k},t) \simeq B(t) k^{-D-1},
\label{eq:Ck}
\ee
where $B(t)$ incorporates the time-dependence. Therefore, one-dimensional energy spectrum is  related to the modal energy as follows:
\be
E_{\rm sp}(k,t) = E({\bf k},t) S_D k^{D-1}  =  B(t) S_D k^{-2},
\ee
where $S_D$ is the area of the $D$-dimensional sphere of unit radius.

Verma et al.~\cite{Verma:PRE2023_coarsening} and Yadav et al. \cite{Yadav:PRE2024} studied the energy transfers in the TDGL and CH equations. Here, we summarize the important results in our previous papers. For delta-correlated initial conditions $\psi({\bf x},0)$, using the Gaussian approximation at early times, Eq.~(\ref{tdgl_tkt}) yields~\cite{Verma:PRE2023_coarsening}  
\be
T({\bf k},t) \simeq -12 \tilde{E}(t) E({\bf k},t) ,
\ee	
where 
\be
\tilde{E}(t) = \sum_{\bf x} \frac{1}{2} |\psi({\bf x},t)|^2.
\ee
Under this approximation,
 \be
\frac{\partial E({\bf k},t)}{\partial t} = k^{2\alpha} \left[2  - 12  \tilde{E}(t) - 2k^2 \right] E({\bf k},t) .
\label{eq:delta_phi}
\ee
We assume that $\tilde{E}(t)$ varies slowly, so the corresponding solution is
\be
E({\bf k},t)   =  \exp \left[ \left(2 - 12 \tilde{E} -2k^2 \right) k^{2\alpha} t \right] E({\bf k},0) .
\label{eq:E_kt_formula}
\ee
 
It is tempting to define the energy flux (as in hydrodynamic turbulence \cite{Kolmogorov:DANS1941Dissipation,Kolmogorov:DANS1941Structure}) for coarsening systems.  Unfortunately, the TDGL and CH equations do not have quadratic invariants, that is, $\sum_{\bf x} |\psi({\bf x},t) |^2$ is not constant in time. Therefore,
\be
\sum_{|{\bf k'}|=0}^\infty  T({\bf k'},t) \ne 0,
\ee
and the energy flux
\be
\Pi({\bf k},t) = \sum_{|{\bf k'}|=0}^k T({\bf k'},t)
\ee
do not represent energy transfers from the large to the small scales \cite{Verma:PRE2023_coarsening}. Nevertheless, the evolution of the modal energy $E({\bf k},t)$ and the transfer function $T({\bf k},t)$ provide valuable information on the coarsening process.  We stress that energy conservation is not necessary to define the energy flux. For example, kinetic energy is not conserved in magneto-hydrodyamic and stably-stratified turbulence. Nevertheless, the kinetic energy flux proves useful for understanding these systems \cite{Verma:JPA2022,Verma:book:ET,Verma:book:BDF, Foldes_2025, Foldes_2024,Zhou_2019, Alexakis2024}. We will explore the energy fluxes in coarsening from this perspective.

 Energy transfers also provide interesting information about the evolution for the biased case where $\psi({\bf x},t=0) = \psi_0 +$small fluctuations ($\psi_0 \neq 0$). In this case, the symmetry between the phases is broken. For the TDGL equation, the average order parameter rapidly evolves to $\pm 1$, whereas the fluctuations about the average are exponentially damped in time. For the CH equation where $\int d{\bf x}~\psi({\bf x},t)$ is conserved, the minority phase forms discrete droplets in a background of the majority phase \cite{Puri:book_edited,Bray:AP1994}. For a detailed discussion of spectral energy transfers in the biased case, we refer the interested reader to \cite{Verma:PRE2023_coarsening,Yadav:PRE2024}.

\subsection{Structure functions}
\label{sub:Sq}

In this subsection, we will compute the $q^{\rm th}$ order structure function defined in Eq.~(\ref{sqpsi}) for the TDGL and CH equations. In the absence of an external field, the system is homogeneous and isotropic, as in hydrodynamic turbulence. Hence, $S_{q}(r,t)$ is a function of $|{\bf r}|=r$. In this subsection, we show that
\begin{eqnarray}
	S_{q}(r,t) = A_q r^{\zeta_{q}} = A_q r,
	\label{eq:zeta_q}
\end{eqnarray}
or $\zeta_q =1$. As we show below, the above simple relation arises due to domain walls, which are analogous to shocks in the Burgers equation \cite{Verma:PA2000}.

We start with the derivation for $D=1$. At late times, both the TDGL and CH equations exhibit a sequence of kinks and anti-kinks, as illustrated in Fig.~\ref{fig:sf1}. We can write $\psi(x,t)$ as the following function in the vicinity of the $i^{\rm th}$ domain wall at $x=\sigma_{i}$: 
\begin{eqnarray}
	\psi(x,t) \simeq \pm \tanh\left(\frac{x-\sigma_i}{\sqrt{2}}\right),
	\label{eq:psi_x}
\end{eqnarray}
where the $+$ and $-$ signs refer to kinks and anti-kinks, respectively. This approximation differs from the exact multi-kink solution in Fig.~\ref{fig:sf1} only in the tails of the kinks and anti-kinks. The difference decays exponentially with the distance between the kinks and becomes asymptotically irrelevant. In addition, the primary contribution to the structure functions comes from the jump region of the kinks. Therefore, the error that arises from this approximation vanishes at late times. The width of the interface in dimensionless units is $\xi = \sqrt{2}$,  while the average distance between adjacent domain walls (interfaces) is denoted by $R(t)$. 

We define the contribution to $S_q(r,t)$ from a single kink as
\begin{eqnarray}
	S_q(r,t) &=& \frac{1}{L}\int_{0}^{L}\left|\psi\left(x + r,t\right) - \psi\left(x,t \right)\right|^{q}dx,
	\label{eq:sftdgl2}
\end{eqnarray}
where $L$ is the size of the system. Assuming that the kinks are spaced far apart at late times, we calculate the structure function in the independent-kink approximation~\cite{Weinan:PRL1999}.

We simplify the notation by setting $\sigma_i = 0$.  For small $r$, which is quantified using $r < \xi$,
\begin{eqnarray}
	\tanh\left({\frac{x + r}{\sqrt{2}}}\right) -\tanh\left(\frac{x}{\sqrt{2}}\right) \simeq \text{sech}^2\left(\frac{x}{\sqrt{2}}\right) \frac{r}{\sqrt{2}}.
	\label{eq:del_psi_small_r}
\end{eqnarray}
Therefore, 
\bea
S_q(r,t) & \simeq &\frac{1}{L} \int_{-\infty}^\infty
\left|\tanh\left( \frac{x + r}{\sqrt{2}} \right) - \tanh\left( \frac{x}{\sqrt{2}} \right)\right|^{q}dx   \nonumber \\
& \simeq & \frac{\sqrt{2}}{L} \left( \frac{r}{\sqrt{2}} \right)^q   
\int_{-\infty}^\infty (\cosh y)^{-2q} dy \nonumber \\
& = & \frac{\sqrt{2}}{L} \left( \frac{r}{\sqrt{2}} \right)^q B(q, 1/2).
\label{eq:Sq1D_r_less}
\eea
Here, $B(q,1/2)$ is the \textit{Beta function}~\cite{Abramowitz:book}:
\be
B(a,b) = \frac{\Gamma (a) \Gamma (b)}{\Gamma (a+b)}.
\ee
The time-dependence arises from the number of defects in the system. For $N_0(t)$ kinks and anti-kinks that are spaced sufficiently apart, the contributions can be added to obtain
\be
S_q(r,t) \simeq \frac{\sqrt{2}}{L} \left( \frac{r}{\sqrt{2}} \right)^q N_0(t) B(q, 1/2).
\label{eq:small_r_scaling}
\ee

Next, we compute $S_q(r,t)$ for $\xi \ll r \ll R(t)$. For a single kink,
when $x$ and $x+r$ are on the same side of the kink (Case A),
\bea
\tanh\left({\frac{x + r}{\sqrt{2}}}\right) -\tanh\left(\frac{x}{\sqrt{2}}\right) \simeq  0.
\eea
However, when $x$ and $x+r$ are on different sides of the kink (Case B),
\bea
\tanh\left({\frac{x + r}{\sqrt{2}}}\right) -\tanh\left(\frac{x}{\sqrt{2}}\right) \simeq 2.
\eea
Therefore, for $\xi \ll r \ll R(t) $, we obtain
\bea
S_q(r,t) &=&\frac{1}{L} \int_{\mathrm{Case~A}}
\left|\tanh\left( \frac{x + r}{\sqrt{2}} \right) - \tanh\left( \frac{x}{\sqrt{2}} \right)\right|^{q}dx  \nonumber \\
&& +
\frac{1}{L} \int_{\mathrm{Case~B}}
\left|\tanh\left( \frac{x + r}{\sqrt{2}} \right) - \tanh\left( \frac{x}{\sqrt{2}} \right)\right|^{q}dx
\nonumber \\
& \simeq & 0 + \lim_{\epsilon \rightarrow 0}~\frac{1}{L} 2^q
\int_{-r+\epsilon}^{-\epsilon} dx \nonumber \\
& \simeq & \frac{r}{L} 2^q.
\label{eq:sftdgl5}
\eea
For $N_0(t)$ kinks and anti-kinks,
\be
S_q(r,t) \simeq  \frac{r}{L}N_0(t) 2^{q}.
\label{eq:Sq1D_r_gtr}
\ee

Next, we derive $S_q(r,t)$ for 2D coarsening systems. We consider a domain of size $R(t)$ within a system of size $L \times L$, as illustrated in Fig.~\ref{fig:sf2}. The system consists of two phases, $\psi >0$ (denoted 
$+$) and $\psi <0 $ (denoted $-$). They are separated by a domain wall whose circumference is $l_c$. In this 2D system,
\begin{eqnarray}
	S_{q}(r,t) = \frac{1}{L^2}\int\int |\psi(\mathbf{x} + \mathbf{r},t) - \psi(\mathbf{x},t)|^q dx dy.
\end{eqnarray}
For small $r < \xi$, $\Delta \psi =\psi(\mathbf{x} + \mathbf{r}, t) - \psi(\mathbf{x}, t))$ for 2D is more complicated than Eq.~(\ref{eq:del_psi_small_r}). However, we can approximate the scaling behavior as
\be
S_q(r,t) \propto \left( \frac{r}{\sqrt{2}} \right)^q \frac{1}{L^2}
\ee
for this regime. However, for $\xi \ll r \ll R(t)$, 
\bea
S_q(r,t) & \simeq & \frac{1}{L^2} \int_C |\psi(\mathbf{x} + \mathbf{r},t) - \psi(\mathbf{x},t)|^q~dn~ds ,
\eea
where ${\bf n}$ denotes the transverse direction and ${\bf s}$ is the contour variable. The transverse integral $\int_{-r+\epsilon}^{-\epsilon} dn$ yields $r$, while the integral along the contour yields $\int ds \simeq l_c$. Hence, in 2D, 
\be
S_q(r,t) \simeq r \frac{2^q l_c }{L^2} .
\label{eq:Sq2D_r_gtr}
\ee
For $N_0(t)$ domains that are spaced sufficiently far apart, i.e., the distance $r$ does not span multiple droplets, $l_c$ simply generalizes to the total perimeter length of defects.

We can consider a limiting case where $\psi$ is hardened \cite{op88,po88}, i.e., the smooth domain walls are replaced by step functions. This is done by assigning $\psi=1$ for $\psi>0$, and $\psi=-1$ for $\psi<0$. For hardened $\psi$, the width of the domain wall shrinks to zero and $\Delta \psi$ is either 0 or 2. For this case,
\be
S_q(r,t) =
\begin{cases}
	r \frac{2^q N_0 }{L}~~~~\text{for 1D}, \\
	r \frac{2^q l_c }{L^2}~~~~~\text{for 2D}.
	\label{eq:Sq_r_gtr}
\end{cases}
\ee
It is straightforward to generalize the above result to 3D:
\be
S_q(r,t) = r \frac{2^q A(t) }{L^3},
\ee
where $A(t)$ is the surface area of the domain boundary.

 In this paper, we compute the above structure functions for TDGL and CH systems.  However, we add a cautionary remark that it is sometimes difficult to extract information on small-scale fluctuations using only the two-point structure function. Cho~\cite{Cho19} proposed a technique based on multi-point structure functions to overcome this difficulty. Fortunately, the two-point structure functions described here capture both small-scale and large-scale properties quite well.

\section{Numerical Results}
\label{sec:NR}

\subsection{Simulation details}

We simulated the TDGL and CH equations in $D=1,2$ using the finite-difference method. The size of the system, $L$ in 1D and $L^2$ in 2D, and their respective grids $N$ and $N^2$ are listed in Table~\ref{tab:sim_params}. The grid spacing
\begin{table}
\caption{Simulation parameters for the CH and TDGL equations.}
\begin{tabular}{ccccccc}
Model & $D$ & $L$ & $N$ & $\Delta x = L/N$ & $\Delta t$ \\
\hline
CH & 1 & 100 & 512 & 0.195 & $10^{-4}$ \\
TDGL & 1 & 100 & 1024 & 0.097 & $10^{-3}$ \\
CH & 2 & 128 & 128 & 1 & $10^{-2}$ \\
TDGL & 2 & 128 & 128 & 1& $10^{-2}$ \\
\end{tabular}
\label{tab:sim_params}
\end{table}
$\Delta x (= L/N)$ and the time step $\Delta t$ are also listed in the table. We employ periodic boundary conditions in all directions. Note that the grid spacing $\Delta x$ is smaller than the domain-boundary width $\xi = \sqrt{2}$. This is important for resolution of the defect. The time step $\Delta t$ satisfies a {\it stability criterion} \cite{Puri:book_chapter}. For the CH equation,
\begin{eqnarray}
	\Delta t \le \frac{(\Delta x)^4}{4D [(\Delta x)^{2} + 2D]} .
\end{eqnarray}
For the TDGL equation, we require the following:
\begin{eqnarray}
	\Delta t \le \frac {(\Delta x)^2}{(\Delta x)^{2} + 2D}.
\end{eqnarray}
For the initial condition $\psi ({\bf x},0)$, we chose random fluctuations of small amplitude with zero mean. This mimics the disordered state prior to the temperature quench.

\begin{figure}[htb]
\centering
\includegraphics[width=0.75\linewidth]{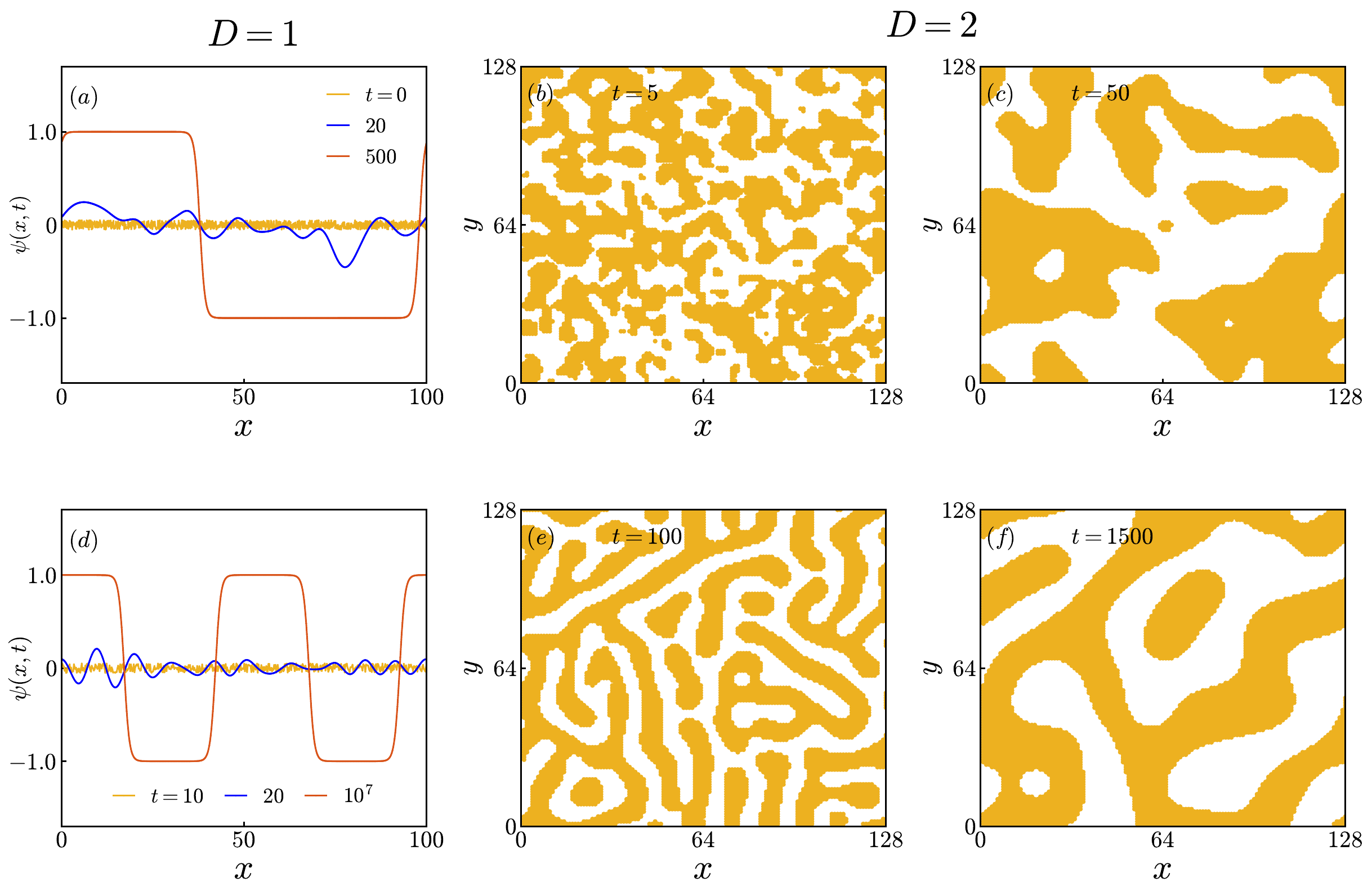}
\caption{Plots showing the evolution of $\psi({\bf x},t)$: (a) For 1D TDGL equation, $\psi(x,t)$ vs. $x$ at $t=0, 20, 500$. (b)-(c) For 2D TDGL equation, density plots of $\psi({\bf x},t)$ at $t=5$ and 50. Points with $\psi > 0$ are marked yellow, and points with $\psi < 0$ are unmarked. (d) For 1D CH equation, $\psi(x,t)$ vs. $x$ at $t=10, 20, 10^7$. (e)-(f) For 2D CH equation, density plots of $\psi({\bf x},t)$ at $t=100$ and 1500.}
\label{fig:profile}
\end{figure}

In Fig.~\ref{fig:profile}, we show the evolution of $\psi({\bf x},t)$ for the TDGL equation (top row) and the CH equation (bottom row). The first column shows the evolution of the 1D system, whereas the second and third columns show the evolution of the 2D system. As seen in the figure, the structures grow with time in both the 1D and 2D cases. For the sake of brevity, we do not present simulation results for energy transfers, etc. here. The interested reader is referred to  \cite{Verma:PRE2023_coarsening,Yadav:PRE2024}, where these quantities are discussed in length.

In the following subsections, we present the structure functions and their scaling for the TDGL and CH equations.

\subsection{Structure functions for the TDGL equation}
\label{sub:TDGL_Sq}

For the TDGL equation, we compute $S_q(r,t)$ at $t=500$ for 1D and $t=100$ for 2D. We averaged across the whole box for each value of $r$. 

\begin{figure}[htb]
\centering
\includegraphics[width=0.7\linewidth]{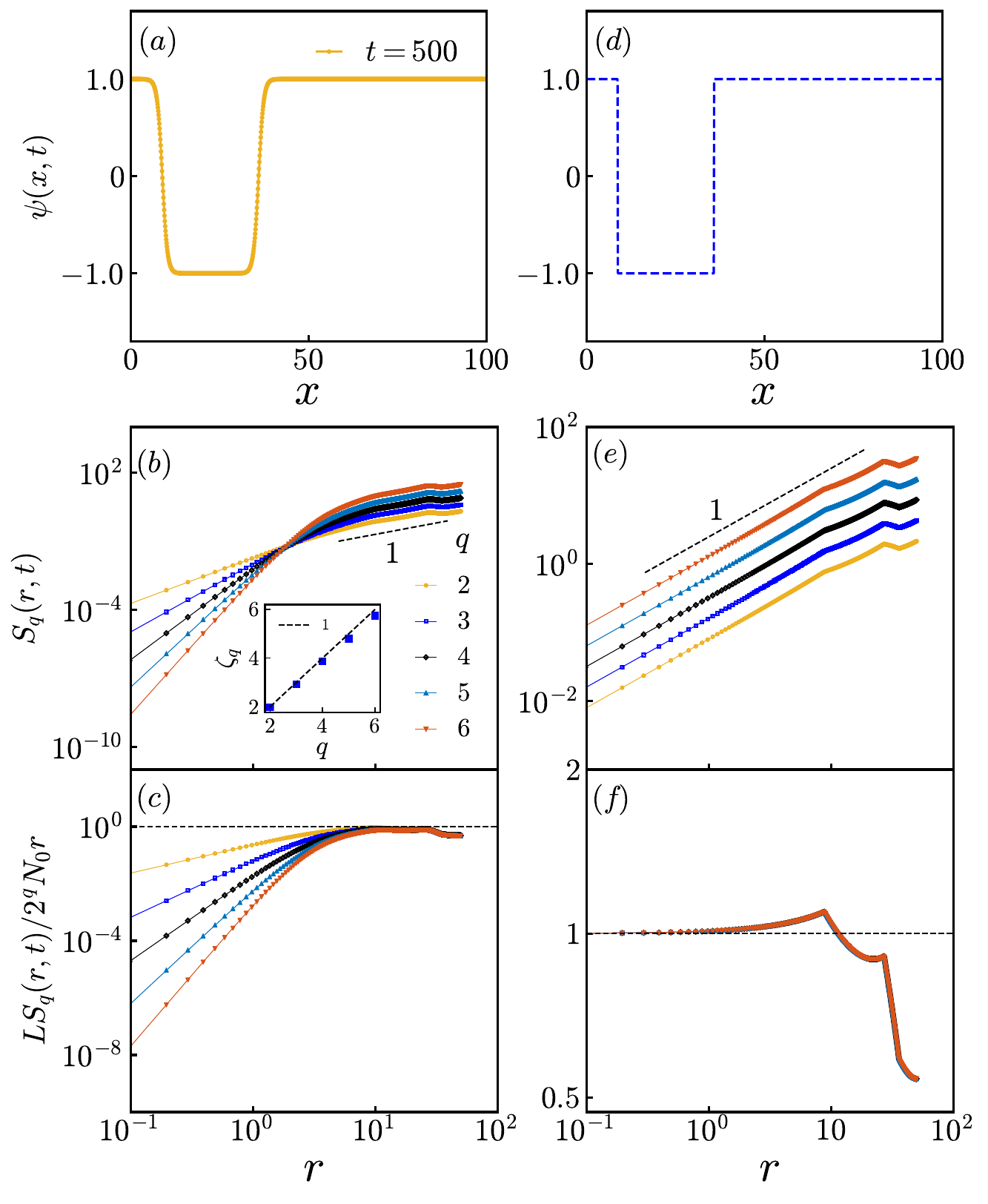}
\caption{1D TDGL equation: (a) Profile $\psi(x,t=500)$ vs. $x$. (b) For the profile in (a), structure function $S_q(r,t)$ vs. $r$ for $q =2$ to $6$. The inset shows $\zeta_q$ vs. $q$ for $r <\xi$. (c) For the data in (b), normalized structure function $L S_q(r,t)/(2^q N_0 r) \simeq 1$ with $N_0=2$. The symbols used are the same as those in (b). (d) Hardened $\psi(x,t=500)$ vs. $x$. (e)-(f) $S_q(r,t)$ and normalized $S_q(r,t)$ for the hardened profile in (d). The symbols used are the same as those in (b).}
\label{fig:sf_TDGL1d}
\end{figure}

Fig.~\ref{fig:sf_TDGL1d} shows $\psi(x,t)$ and its associated structure functions for the 1D TDGL equation. The snapshot in Fig.~\ref{fig:sf_TDGL1d}(a) shows a state with a kink and anti-kink pair. Fig.~\ref{fig:sf_TDGL1d}(b) plots $S_q(r,t)$ vs. $r$ for $q=2$ to 6. As shown in the frame, $S_q(r,t) \sim r^q$ for $r < \xi = \sqrt{2}$ and $S_q(r,t) \sim r$ for $r > \xi$.  To verify the analytical prediction for the small-$r$ regime, we extract the corresponding scaling exponent $\zeta_q$ from the power-law behavior in Fig.~\ref{fig:sf_TDGL1d}(b). We plot $\zeta_q$ as a function of $q$ in the inset of Fig.~\ref{fig:sf_TDGL1d}(b), and find excellent agreement with the analytical prediction in Eq.~\eqref{eq:Sq1D_r_less}. In Fig.~\ref{fig:sf_TDGL1d}(c), we show the normalized quantity $L S_q(r,t)/(2^q N_0 r) \simeq 1$ for $r \gg \xi$ and $N_0=2$. These results are consistent with Eq.~(\ref{eq:Sq1D_r_gtr}).  Also note that $S_q(r,t) \propto r$ scaling is valid for $r < 10$, which is approximately the distance between the kink and anti-kink. This is consistent with our earlier discussion.

As mentioned earlier, the hardened $\psi(x,t)$ is often used in the coarsening literature to illustrate the Porod tail \cite{op88,po88}. Fig.~\ref{fig:sf_TDGL1d}(d) shows the hardened $\psi(x,t)$ for the profile of Fig.~\ref{fig:sf_TDGL1d}(a). Note the step function jumps in the figure. Figs.~\ref{fig:sf_TDGL1d}(e), (f) show $S_q(r,t)$ and its normalized counterpart for hardened data. The hardened data do not show the $r^q$ behavior for small $r$ because the interfacial width has been set to 0 by hardening.

\begin{figure}[htb]
\centering
\includegraphics[width=0.75\linewidth]{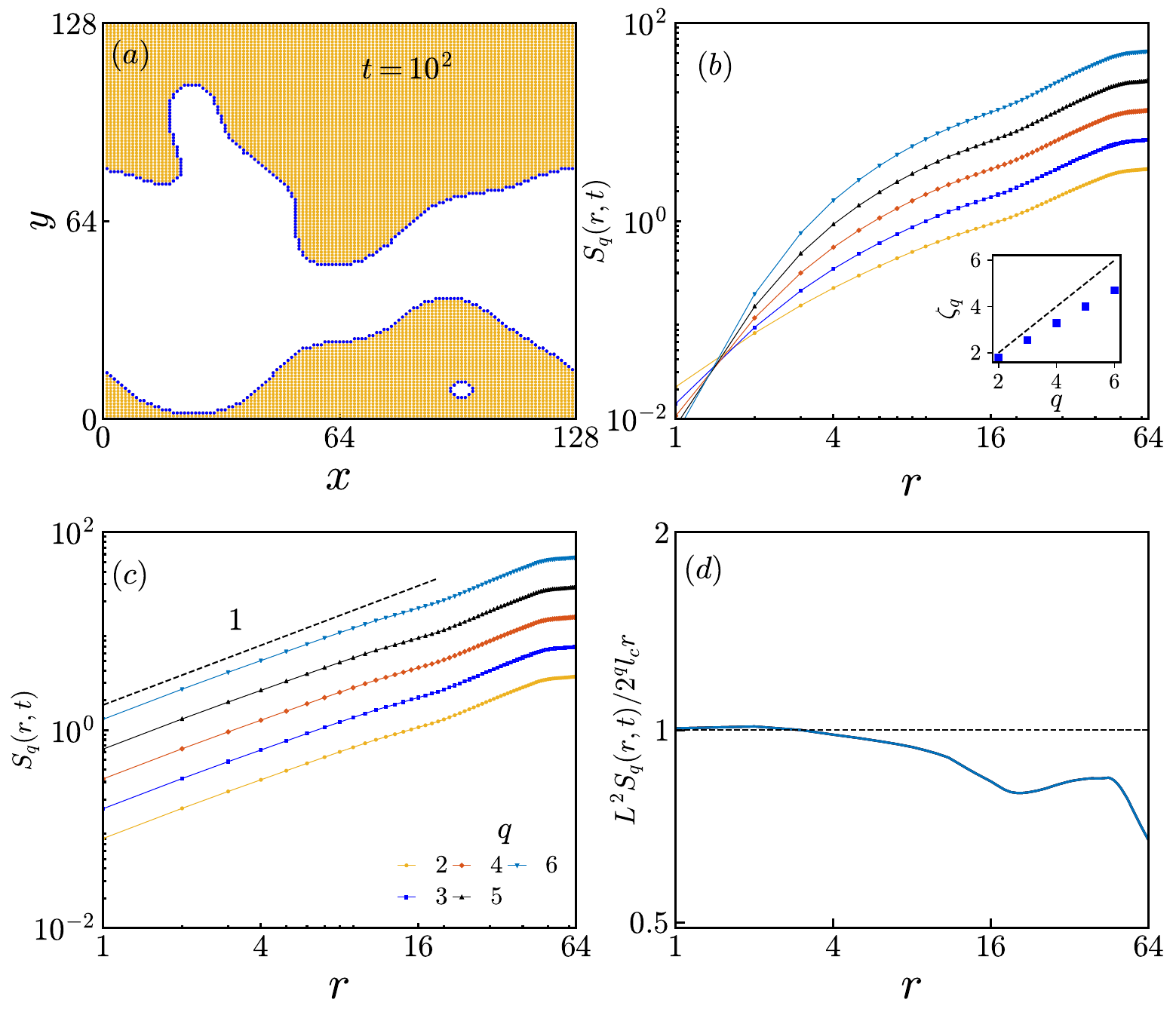}
\caption{2D TDGL equation: (a) Density plot of the hardened field $\psi({\bf x},t=100)$. 
(b) Structure function $S_q(r,t)$ vs.\ $r$ for $q=2$ to $6$. The inset shows $\zeta_q$ vs. $q$ for $r <\xi$. The deviation from $\zeta_q = q$ is due to under-resolution of the interface region.
(c) Structure function $S_q(r,t)$ vs.\ $r$ for $q=2$ to $6$ computed from the hardened data shown in (a). 
(d) For the data in (c), the normalized structure function $L^2 S_q(r,t)/(2^q l_c r) \simeq 1$. 
The symbols used are the same as those in (c).}

\label{fig:sf_TDGL2D}
\end{figure}

Next, we compute $S_q(r,t)$ for the 2D TDGL equation. In Fig.~\ref{fig:sf_TDGL2D}(a), we present a snapshot of $\psi({\bf x},t)$ at 
$t=100$; here the yellow region has $ \psi > 0 $, while the white region has $ \psi < 0 $. The blue dots represent the domain boundary. To compute the circumference of the domain boundary, we scan the system to identify points where the field $\psi({\bf x}, t)$ changes sign from $1 \to -1$ or vice versa. First, we perform a horizontal scan, detecting the points where $\psi(x,y,t)~\psi(x+\Delta x,y,t) < 0$. Second, we perform a scan in the $y$-direction, identifying the points where $\psi(x,y,t)~\psi(x,y+\Delta x,t) < 0$.  A combination of horizontal and vertical scans ensures that the entire domain boundary is captured. We remove duplicate entries to avoid double counting. This procedure yields $l_c$, the circumference of the boundary.

 We use both the original and the hardened versions of $\psi({\bf x},t)$ to compute the structure functions $S_q(r,t)$ for $q=2$ to $6$, and present the results in Fig.~\ref{fig:sf_TDGL2D}(b)–(c) respectively. The interface region contains only a couple of points for the simulation mesh sizes we use. Therefore, we do not obtain an accurate estimate of $\zeta_q$. We plot $\zeta_q$ vs. $q$ as an inset of Fig.~\ref{fig:sf_TDGL2D}(b), and see that $\zeta_q$ increases linearly with $q$. In Fig.~\ref{fig:sf_TDGL2D}(d), we plot the normalized structure function for the data in Fig.~\ref{fig:sf_TDGL2D}(c).

\subsection{Structure functions for the CH equation}

\begin{figure}[htb]
\centering
\includegraphics[width=0.6\linewidth]{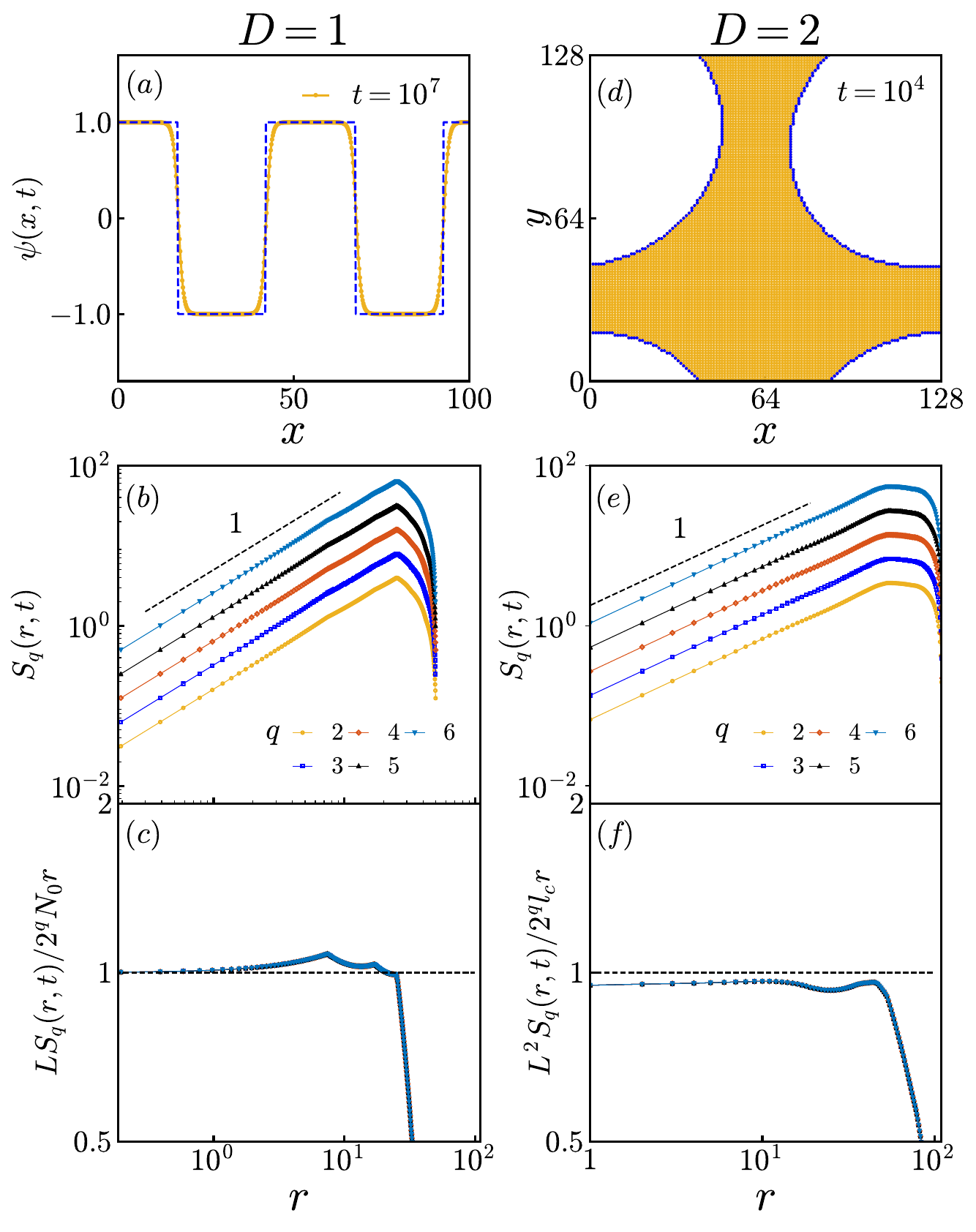}
\caption{1D and 2D CH equations: (a) For 1D CH equation, $\psi(x,t=10^7)$ vs. $x$ and its hardened counterpart. (b) Structure function $S_q(r,t)$ for  hardened profile in (a). (c) For data in (b), normalized structure function $L S_q(r,t)/(2^q N_0 r) \simeq 1$. (d) For 2D CH equation, density plot of hardened $\psi({\bf x},t=10^4)$. (e) $S_q(r,t)$ vs. $r$ for order parameter field in (d). (f) Normalized structure function $L^2 S_q(r,t)/(2^q l_c r) \simeq 1$ for data in (e).}
\label{fig:CH}
\end{figure}

In this subsection, we present the structure functions for the CH equation in 1D and 2D. In Fig.~\ref{fig:CH}(a), we show the 1D profile of $\psi(x,t)$ at $t=10^7$. The hardened profile is shown in Fig.~\ref{fig:CH}(a) using a dashed line. In Fig.~\ref{fig:CH}(b), we exhibit the structure functions $S_q(r,t)$ for the hardened $\psi(x,t)$ with $q$ from 2 to 6.  As shown in the figure, $S_q(r,t) \propto r$. We verify the scaling in Eq.~(\ref{eq:Sq1D_r_gtr}) by showing in Fig.~\ref{fig:CH}(c) that 
\be 
\frac{L S_q(r,t)}{2^q N_0 r} \simeq 1,
\ee
with $N_0=4$ (2 kinks and 2 anti-kinks in Fig.~\ref{fig:CH}(a)).

\begin{figure}[htb]
\centering
\includegraphics[width=0.7\linewidth]{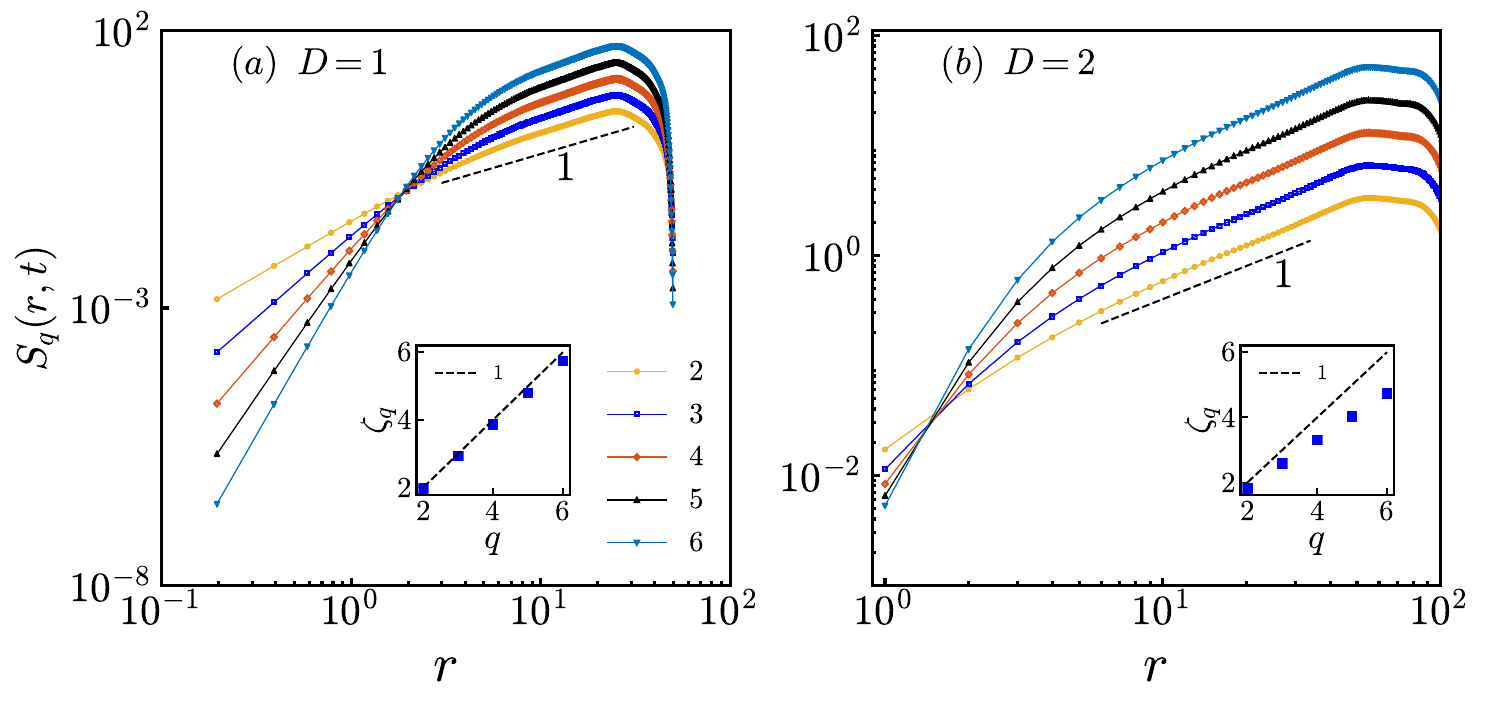}
\caption{1D and 2D CH equations: Structure function $S_q(r,t)$ vs. $r$ for the unhardened profile in Fig.~\ref{fig:CH}(a). The inset shows $\zeta_q$ vs. $q$. (b) $S_q(r,t)$ vs. $r$ for the unhardened order parameter field in Fig.~\ref{fig:CH}(d). The inset shows $\zeta_q$ vs. $q$.}
\label{fig:CH2}
\end{figure}

\begin{figure}[htb]
\centering
\includegraphics[width=0.7\linewidth]{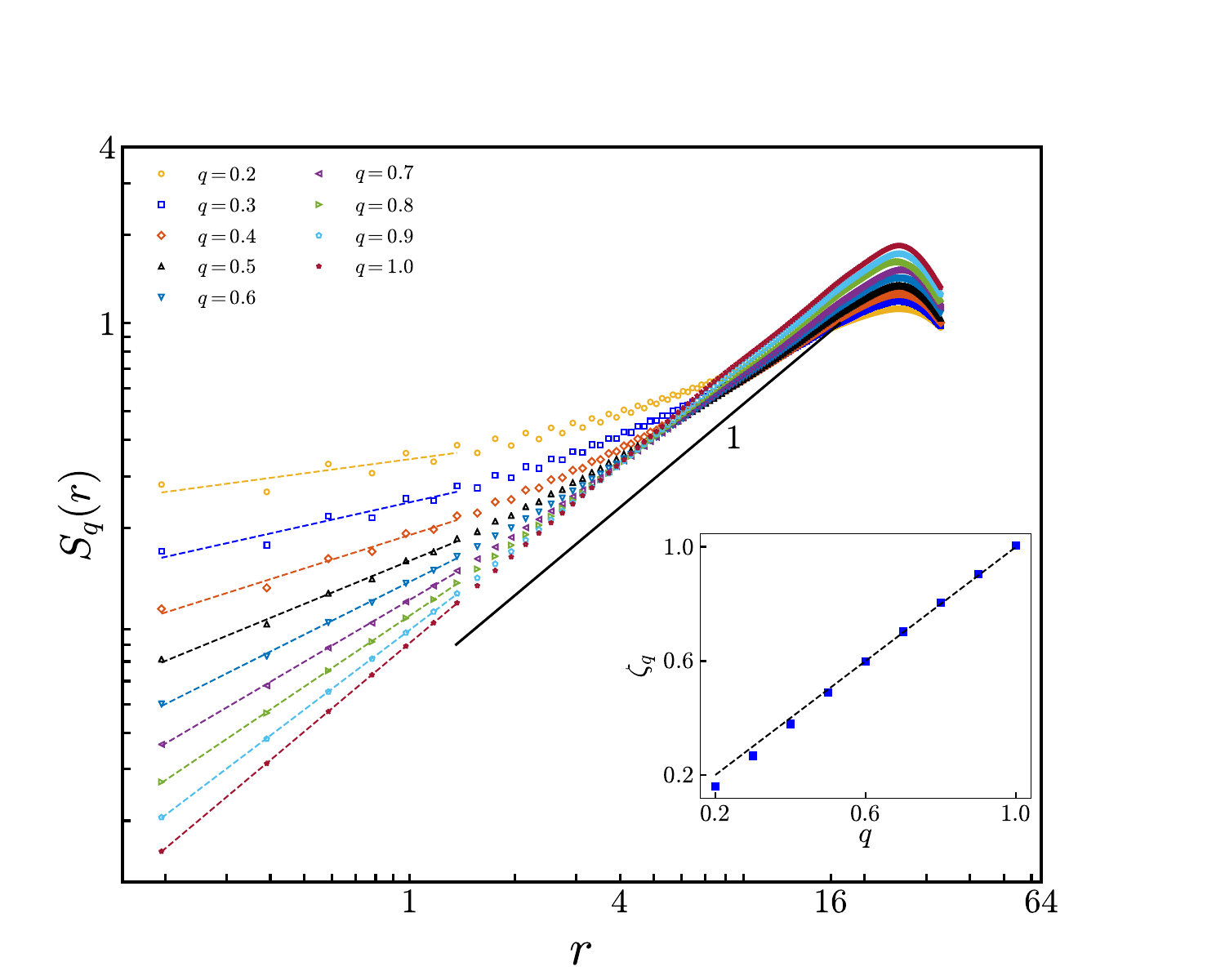}
\caption{1D CH equation: $S_q(r,t)$ vs. $r$ for the unhardened profile in Fig.~\ref{fig:CH}(a) for $q \leq 1$. The inset shows $\zeta_q$ for $r < \xi$.}
\label{fig:CH1d_q_lt1}
\end{figure}
For the 2D CH equation, we obtain the structure functions at $t=10^4$. In Fig.~\ref{fig:CH}(d), we show a density plot of the hardened  $\psi({\bf x},t)$. The brown region represents $\psi = 1$, while the white region represents $\psi = -1$. The blue dots show the domain boundary between $\psi=1$ and $-1$. The structure functions for this profile are shown in Fig.~\ref{fig:CH}(e). In Fig.~\ref{fig:CH}(f), we show that $L^2 S_q(r,t)/(2^q l_c r) \simeq 1$, where $l_c$ is the net circumference of the boundary in Fig.~\ref{fig:CH}(d).

 To clarify the small-$r$ behavior, Fig.~\ref{fig:CH2}(a)–(b) plots $S_q(r,t)$ vs. $r$ for the original profiles shown in Figs.~\ref{fig:CH}(a) and (d). We plot $\zeta_q$ vs. $q$ for the 1D data as an inset of Fig.~\ref{fig:CH2}(a), revealing the expected $\zeta_q = q$ behavior. We repeated this analysis for $D=2$; however, the results underestimate $\zeta_q = q$ as for the $D=2$ TDGL equation.

 Before concluding, we stress that the definition of structure functions is not restricted to integer $q$. In Fig.~\ref{fig:CH1d_q_lt1}, we show $S_q$ vs. $r$ for the unhardened profile of the $D=1$ CH equation in Fig.~\ref{fig:CH}(a) for $0< q \le 1$. The results are consistent with Eq.~\eqref{eq:small_r_scaling} for $r < \xi$. The exponent $\zeta_q$ for small $r$ is extracted from the corresponding power-law behavior of $S_q(r,t)$. In the inset of Fig.~\ref{fig:CH1d_q_lt1}, we confirm $\zeta_q \sim q$. For $r > \xi$, $\zeta_q \sim 1$, as expected.

\section{Conclusions}
\label{sec:conclusion}

Coarsening and turbulence are two important nonequilibrium phenomena that have received great attention in the literature. In this and earlier papers~\cite{Verma:PRE2023_coarsening,Yadav:PRE2024}, we have discussed how tools developed in the context of turbulence can be gainfully applied to understand domain growth problems. In coarsening, the segregation of two phases via an interface is similar to the separation of fast and slow moving particles by a shock in Burgers equation. In this paper, we show that the structure functions of the time-dependent Ginzburg-Landau (TDGL) and Cahn-Hilliard (CH) equations (which describe coarsening) are analogous to the structure functions of Burgers equation (which is a standard model of turbulence). In particular,  we show that for the TDGL and CH equations, the $q^{\rm th}$ order structure function $S_q(r,t) \propto r^q$ for $r < \xi$ and $S_q(r,t) \propto r$ for $r > \xi$. Here, $\xi$ is the interfacial width. These relations hold for all dimensions but the pre-factors depend on the dimensionality. Since $S_q(r) \propto r$ for large $r$, we conclude that the TDGL and CH equations exhibit {\it intermittency}, in analogy with turbulent systems. Using numerical simulations, we verified the above scaling relations inclusive of pre-factors.

 In hydrodynamic and Burgers turbulence, the kinetic energy cascades from large scales to intermediate and small scales. The structure of nonlinearity and conservation laws in domain growth problems are somewhat different from those of hydrodynamic turbulence. Nevertheless, the nonlinear energy transfers in coarsening systems yield valuable insights, as shown in recent works~\cite{Verma:PRE2023_coarsening,Yadav:PRE2024}. In these works, the nonlinear term induces an effective diffusion, which matches with the energy injection term of the system (the first term on the right-hand-side of Eq.~\eqref{eq:en}).  In the present work, we revisit these issues and relate them to the scaling of coarsening lengths. We also remark that the coarsening is not caused by an inverse cascade of $\psi^2$; instead, it occurs due to additional diffusion induced by nonlinearity.

We expect this framework of energy transfers and structure functions, developed in analogy with turbulence studies, to be relevant in other coarsening systems as well. An important system in this context is {\it Model H} \cite{Hohenberg:RMP1977}, which models the phase separation kinetics of a binary fluid mixture. Model H is more complicated than the equations discussed here as it describes the coupled dynamics of the order parameter and the fluid velocity fields. The energy transfers in Model H are more complex due to the advection of the velocity field~\cite{Bratanov:PNAS2015}. Another fascinating class of pattern-forming systems (e.g., chemical reactions, animal skin patterns, population dynamics, etc.) is described by {\it reaction-diffusion systems} \cite{Cross:RMP1993,Cross:book:Pattern}. Much of the past work on these systems has focused on instabilities and the selection of length scales. We believe that nonlinear energy transfers should play an important role in understanding pattern formation in these systems. We will explore some of these directions in the future.

In conclusion, tools taken from turbulence studies (e.g. energy transfers, nonlinear dissipation, structure factors, etc.) offer a fresh perspective on domain growth problems. More generally, we believe that these techniques can be gainfully applied in diverse examples of pattern formation, which are ubiquitous in scientific and social disciplines. We hope that our current work will provide an impetus in these directions.

\dataccess{The data supporting this study are available in Zenodo at \url{https://doi.org/10.5281/zenodo.17543981}.}

\funding{This work was partially supported by the J. C. Bose Fellowship awarded to Mahendra Verma.}

\conflict{The authors have no conflict of interest to declare.}

\ack{The authors thank K.R. Sreenivasan and Jörg Schumacher for valuable comments and suggestions and a careful reading of the manuscript. Part of this work was supported by the Anusandhan National Research Foundation, India (Grant Nos. SERB/PHY/2021522 and SERB/PHY/2021473), and the J. C. Bose Fellowship (SERB /PHY/2023488).}

%\disclaimer{Insert disclaimer text here if applicable.}

%%%%%%%%%% Insert bibliography here %%%%%%%%%%%%%%

\bibliographystyle{RS}
\bibliography{bib/journal,bib/book, bib/book_chapter, bib/book_edited} 

\begin{thebibliography}{99}

\bibitem{Manneville:book:Instabilities}
Manneville P. 2004 {\em {Instabilities, Chaos and Turbulence}}.
London: Imperial College Press.

\bibitem{Manneville:book:Structures}
Manneville P. 2014 {\em {Dissipative Structures and Weak Turbulence}}.
San Diego: Academic Press.

\bibitem{Frisch:book}
Frisch U. 1995 {\em {Turbulence: The Legacy of A. N. Kolmogorov}}.
Cambridge: Cambridge University Press.

\bibitem{Kolmogorov:DANS1941Dissipation}
Kolmogorov AN. 1941a  {Dissipation of Energy in Locally Isotropic Turbulence}. {\em Dokl Acad Nauk SSSR} \textbf{32}, 16--18.

\bibitem{Kolmogorov:DANS1941Structure}
Kolmogorov AN. 1941b  {The local structure of turbulence in incompressible viscous fluid for very large Reynolds numbers}. {\em Dokl Acad Nauk SSSR} \textbf{30}, 301--305.

\bibitem{Lesieur:book:Turbulence}
Lesieur M. 2008 {\em {Turbulence in Fluids}}.
Dordrecht: Springer-Verlag.

\bibitem{Frisch:JFM1978}
Frisch U, Sulem PL, Nelkin M. 1978  {A simple dynamical model of intermittent fully developed turbulence}. {\em J. Fluid Mech.} \textbf{87}, 719--736.

\bibitem{Kolmogorov:JFM1963}
Kolmogorov AN. 1962  {A refinement of previous hypotheses concerning the local structure of turbulence in a viscous incompressible fluid at high Reynolds number}. {\em J. Fluid Mech.} \textbf{13}, 82--85.

\bibitem{Meneveau:PRL1987}
Meneveau C, Sreenivasan KR. 1987  {Simple multifractal cascade model for fully developed turbulence}. {\em Phys. Rev. Lett.} \textbf{59}, 1424--1427.

\bibitem{She:PRL1994}
She ZS, Leveque E. 1994  {Universal scaling laws in fully developed turbulence}. {\em Phys. Rev. Lett.} \textbf{72}, 336--339.

\bibitem{Puri:book_edited}
Puri S, Wadhawan V, editors. 2009 {\em {Kinetics of Phase Transitions}}.
Boca Raton, FL: CRC Press.

\bibitem{Bray:AP1994}
Bray AJ. 1994  {Theory of phase-ordering kinetics}. {\em Adv. Phys.} \textbf{43}, 357--459.

\bibitem{Hohenberg:RMP1977}
Hohenberg PC, Halperin BI. 1977  {Theory of dynamic critical phenomena}. {\em Rev. Mod. Phys.} \textbf{49}, 435--479.

\bibitem{Verma:PRE2023_coarsening}
Verma MK, Agrawal R, Yadav PK, Puri S. 2023  {Nonlinear energy dissipation and transfers in coarsening systems}. {\em Phys. Rev. E} \textbf{107}, 034207.
(\href{http://dx.doi.org/10.1103/physreve.107.034207}{10.1103/physreve.107.034207})

\bibitem{Yadav:PRE2024}
Yadav PK, Verma MK, Puri S. 2024  Spectral energy transfers in domain growth problems. {\em Phys. Rev. E} \textbf{110}, 044130.
(\href{http://dx.doi.org/10.1103/physreve.110.044130}{10.1103/physreve.110.044130})

\bibitem{gp82}
Porod G, Glatter O, Kratky O, editors. 1982 {\em \textit{Small-Angle X-Ray Scattering}}.
New York: Academic Press.

\bibitem{opmpl}
Oono Y, Puri S. 1988  Large wave number features of form factors for phase transition kinetics. {\em Mod. Phys. Lett. B} \textbf{02}, 861--867.
(\href{http://dx.doi.org/10.1142/S0217984988000606}{10.1142/S0217984988000606})

\bibitem{Bouchaud:PRE1995}
Bouchaud JP, M\'ezard M, Parisi G. 1995  Scaling and intermittency in Burgers turbulence. {\em Phys. Rev. E} \textbf{52}, 3656--3674.
(\href{http://dx.doi.org/10.1103/PhysRevE.52.3656}{10.1103/PhysRevE.52.3656})

\bibitem{Verma:PA2000}
Verma MK. 2000  {Intermittency exponents and energy spectrum of the Burgers and KPZ equations with correlated noise}. {\em Phys. A: Stat. Mech. Appl.} \textbf{277}, 359--388.

\bibitem{Porod:book_chapter}
Porod G.
p.~35.
In , p.~35. Academic Press, New York, 1982.

\bibitem{Verma:JPA2022}
Verma MK. 2022  {Variable energy flux in turbulence}. {\em J. Phys. A: Math. Theor.} \textbf{55}, 013002.
(\href{http://dx.doi.org/10.1088/1751-8121/ac354e}{10.1088/1751-8121/ac354e})

\bibitem{Verma:book:ET}
Verma MK. 2019 {\em Energy transfers in Fluid Flows: Multiscale and Spectral Perspectives}.
Cambridge: Cambridge University Press.

\bibitem{Verma:book:BDF}
Verma MK. 2018 {\em Physics of Buoyant Flows: From Instabilities to Turbulence}.
Singapore: World Scientific.

\bibitem{Foldes_2025}
{Foldes} R, {Marino} R, {Cerri} SS, {Camporeale} E. 2025  {Characterization of local energy transfer in large-scale intermittent stratified turbulent flows via coarse-graining}. {\em Phys. Rev. Fluids} \textbf{10}, 043803.
(\href{http://dx.doi.org/10.1103/PhysRevFluids.10.043803}{10.1103/PhysRevFluids.10.043803})

\bibitem{Foldes_2024}
{Foldes} R, {Cerri} SS, {Marino} R, {Camporeale} E. 2024  {Evidence of dual energy transfer driven by magnetic reconnection at subion scales}. {\em \pre} \textbf{110}, 055207.
(\href{http://dx.doi.org/10.1103/PhysRevE.110.055207}{10.1103/PhysRevE.110.055207})

\bibitem{Zhou_2019}
{Zhou} M, {Bhat} P, {Loureiro} NF, {Uzdensky} DA. 2019  {Magnetic island merger as a mechanism for inverse magnetic energy transfer}. {\em Phys. Rev. Research} \textbf{1}, 012004.
(\href{http://dx.doi.org/10.1103/PhysRevResearch.1.012004}{10.1103/PhysRevResearch.1.012004})

\bibitem{Alexakis2024}
Alexakis A, Marino R, Mininni PD, van Kan A, Foldes R, Feraco F. 2024  Large-scale self-organization in dry turbulent atmospheres. {\em Science} \textbf{383}, 1005--1009.
(\href{http://dx.doi.org/10.1126/science.adg8269}{10.1126/science.adg8269})

\bibitem{Weinan:PRL1999}
E W, Vanden~Eijnden E. 1999  Asymptotic Theory for the Probability Density Functions in Burgers Turbulence. {\em Phys. Rev. Lett.} \textbf{83}, 2572--2575.
(\href{http://dx.doi.org/10.1103/PhysRevLett.83.2572}{10.1103/PhysRevLett.83.2572})

\bibitem{Abramowitz:book}
Abramowitz M, Stegun IA. 1965 {\em {Handbook of Mathematical Functions}}.
With Formulas, Graphs, and Mathematical Tables. Courier Corporation.

\bibitem{op88}
Oono Y, Puri S. 1988  Study of phase-separation dynamics by use of cell dynamical systems. I. Modeling. {\em Phys. Rev. A} \textbf{38}, 434--453.
(\href{http://dx.doi.org/10.1103/PhysRevA.38.434}{10.1103/PhysRevA.38.434})

\bibitem{po88}
Puri S, Oono Y. 1988  Study of phase-separation dynamics by use of cell dynamical systems. II. Two-dimensional demonstrations. {\em Phys. Rev. A} \textbf{38}, 1542--1565.
(\href{http://dx.doi.org/10.1103/PhysRevA.38.1542}{10.1103/PhysRevA.38.1542})

\bibitem{Cho19}
{Cho} J. 2019  {A Technique for Removing Large-scale Variations in Regularly and Irregularly Spaced Data}. {\em \apj} \textbf{874}, 75.
(\href{http://dx.doi.org/10.3847/1538-4357/ab06f3}{10.3847/1538-4357/ab06f3})

\bibitem{Puri:book_chapter}
Puri S. 2011  Kinetics of Phase Transitions: Numerical Techniques and Simulations. In {\em Computational Statistical Physics: Lecture Notes, Guwahati SERC School} ,  pp. 123--160. Springer.

\bibitem{Bratanov:PNAS2015}
Bratanov V, Jenko F, Frey E. 2015  {New class of turbulence in active fluids}. {\em PNAS} \textbf{112}, 15048--15053.
(\href{http://dx.doi.org/10.1073/pnas.1509304112}{10.1073/pnas.1509304112})

\bibitem{Cross:RMP1993}
Cross MC, Hohenberg PC. 1993  {Pattern formation outside of equilibrium}. {\em Rev. Mod. Phys.} \textbf{65}, 851--1112.

\bibitem{Cross:book:Pattern}
Cross M, Greenside H. 2009 {\em {Pattern formation and dynamics in nonequilibrium systems}}.
Cambridge: Cambridge University Press.

\end{thebibliography}

\end{document}